# Fast Intercalation of Lithium in Semi-Metallic γ-GeSe Nanosheet: A New Group-IV Monochalcogenide for Lithium-Ion Battery Application

Zheng Shu, [a] Xiangyue Cui, [a] Bowen Wang, [a] Hejin Yan, [a] Prof. Yongqing Cai[a]*

[a]Joint Key Laboratory of the Ministry of Education, Institute of Applied Physics and Materials Engineering, University of Macau, Taipa, Macau, China

* Authors to whom any correspondence should be addressed.

E-mail: yongqingcai@um.edu.mo

**Table of Contents**

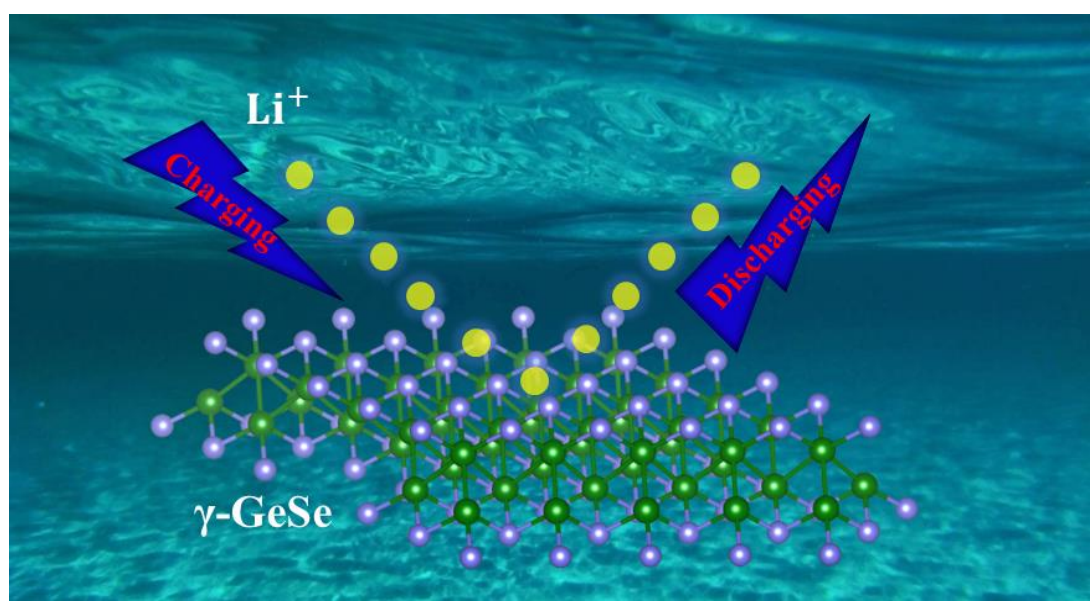

A recently newly synthesized monochalcogenide, γ-GeSe, is demonstrated for potential application of Li exfoliation and lithium-ion battery with the Li species showing a small diffusion barrier of 0.21 eV and the voltage ranging from 0.071-0.015 V. The theoretical storage capacity of γ-GeSe is over 530.36 mAh g$^{-1}$. Such results suggest the γ-GeSe nanosheet can be a promising anode material for lithium-ion battery.

## Abstract

Existence of van der Waals gaps renders two-dimensional (2D) materials ideal passages of lithium for being used as anode materials. However, the requirement of good conductivity significantly limits the choice of 2D candidates. So far only graphite is satisfying due to its relatively high conductivity. Recently, a new polymorph of layered germanium selenide (γ-GeSe) was proven to be semimetal in its bulk phase with a higher conductivity than graphite while its monolayer behaves semiconducting. In this work, by using first-principles calculations, we examined the possibility of using this new group-IV monochalcogenide, γ-GeSe, as anode in the Li-ion battery (LIBs). Our studies revealed that Li atom would form an ionic adsorption with adjacent selenium atoms at the hollow site and exist in cationic state (lost 0.89 e to γ-GeSe). Results of climbing image-nudged elastic band show the diffusion barrier of Li is 0.21 eV in the monolayer limit, which can activate a relatively fast diffusion even at room temperature on the γ-GeSe surface. The calculated theoretical average voltages range from 0.071 to 0.015 V at different stoichiometry of Li$_x$GeSe with minor volume variation, suggesting its potential application as anode of LIBs. The predicted moderate binding energy, a low open circuit voltage (comparable to graphite) and a fast motion of Li suggests that γ-GeSe nanosheet can be chemically exfoliated via Li intercalation and a promising candidate as the anode material for LIBs.

## Introduction

To date, rechargeable lithium-ion batteries (LIBs) have formed a crucial part in human daily life,[1, 2] especially in portable electronic devices and electric vehicles. Hence, LIBs brought about significant scientific and technological revolution to the humanity, and the 2019 Nobel Prize in Chemistry was jointly awarded to three scientists for their groundbreaking contributions of the LIBs development. However, the rapid development of current high-tech applications requires faster charge/discharge rate, longer cycle life, more reliable safety and higher storage capacity of LIBs.[3] Intercalated Li atoms are oxidized on the anodic (negative) electrode and move towards the cathodic (positive) electrode during the discharge process,[4] therefore electrode materials are the key components for accommodating Li species in LIBs. Commercial host for Li-ion intercalation is graphite as anodic (negative) electrode. In the past decade, significant efforts have been devoted to exploring other electrode materials for metal-ion battery, especially for LIBs.[5-14] Two-dimensional (2D) materials are considered as promising energy storage materials due to their large surface area and flexible mechanical properties. For instance, Hardikar et al. achieved the diffusion barriers for LIBs of 0.29, 0.49 and 0.35 eV on graphene, mono-vacancy graphene and di-vacancy graphene, respectively.[15] A newly popgraphene was presented by Wang et al. with a low diffusion barrier of about 0.3 eV.[16] Çakır et al performed first-principles calculations for Li adsorption on the $Mo_2C$ monolayer and predicted a theoretical capacity of over 400 mA h $g^{-1}$.[17, 18] Mxenes are widely used as anode materials in experiments.[19, 20] Exploring novel 2D materials for LIBs is an ever-growing field.

Recently, the family of group-IV monochalcogenides MX (M = Ge, Sn, X = S, Se, Te) with unique "puckered" layered structures has been layered materials in vogue.[21-30] The group-IV monochalcogenides are usually analogues of phosphorene and have gained extensive research interests as ferroelectric,[31-34] thermoelectric,[35-37] photocatalysis[38, 39] and electro-catalysis[40, 41]. The α and β phases with orthorhombic crystal lattice are the normal polymorphs of group-IV monochalcogenides which are most well-studied. However, a recent work[42] reported that a new layered structure (γ-phase) of group-IV monochalcogenides can exist in stabilized state and are energetically favorable compared to α and β phases. Interestingly, the γ phase of germanium selenide (GeSe) has been recently successfully synthesized on the h-BN substrate[43] through chemical vapor deposition (CVD). The γ-GeSe was reported to be semimetal in bulk phase and has a high electrical conductivity of $3\times10^5$ S/m which is even higher than graphite. This intriguing feature warrants a study on its promising applications in LIBs.

In this work, we investigated the adsorption and diffusion behaviors of Li through the surface or van der Waals (vdW) gaps of γ-GeSe surface by using first-principles calculations. According to the calculated results, Li atom is able to be absorbed at γ-GeSe surface with an adsorption energy of -1.68 eV, forming a strong interaction between Li and γ-GeSe. Moreover, the diffusion barrier and voltage are 0.21 eV and 0.071-0.015 V, respectively. Our reported energetics and kinetics of Li species in γ-GeSe would be useful for guiding experiments for interlayer intercalation for exfoliation and fabricating promising LIBs.

## Computational Details

All structural optimizations and energy calculations through the first-principles framework were performed in the Vienna *Ab initio* Simulation Package (VASP).[44] The interaction between core and valence electrons was treated by projected augmented wave (PAW)[45] potential. Spin polarized generalized gradient approximation (GGA) with Perdew—Burke—Ernzerhof (PBE)[46] pseudopotential were used to describe the exchange-correlation functional. A supercell with 4 × 4 primitive cells of γ-GeSe {$Ge_{64}Se_{64}$} was chosen for Li adsorption and diffusion, which results in Li/substrate ratio of 0.016. The cutoff energy for the plane-wave basis was set to 400 eV, and Gaussian smearing was used with a SIGMA value 0.05. The convergence criteria of atomic relaxation were set to 0.01 meV for electronic loops and 0.01 eV Å$^{-1}$ on each atom for ionic loops, respectively. For the γ-GeSe nanosheet, a vacuum layer of 20 Å was placed to avoid mirror interlayer interaction between periodic images. DFT-D3[47] approach of Grimme's semiempirical, which was important for Li adsorption,[11] was taken into account to accurately describe the vdW interaction between lithium atoms and γ-GeSe nanosheet. Γ-centered 3 × 3 × 1 *k*-point grids for γ-GeSe nanosheet and 3 × 3 × 3 for γ-GeSe bulk phase using Monkhorst-Pack scheme were adopted, and a denser 6 × 6 × 1 *k*-point grids were used to calculate the band structure and density of states (DOS). Bader charge analysis[48] was carried out to study the charge transfer between Li and substrate. To figure out the diffusion energy barriers, climbed nudged elastic band (CI-NEB)[49] method were performed. A supercell of 2 × 2 × 1 was utilized to evaluate the specific capacity of γ-GeSe. To evaluate the kinetic stability of monolayer γ-GeSe, a large supercell of 7 × 7 × 1 is used for the calculation of phonon dispersion based on the finite displacement method[50]. *Ab initio* molecular dynamics (AIMD)[51] simulations were carried out for the thermodynamics stability evaluation at room temperature (300 K) with a duration of 10 ps. VASPKIT[52] and VESTA programs are used as the post-processing tools for data analysis and visualization.

## Results and Discussion

**Energetics of Li on γ-GeSe.**

The optimized crystal structure of γ-GeSe nanosheet with a double-layer honeycomb lattice is shown in Figure 1a, where four atomic layers (Se-Ge-Ge-Se) are located at a A-B-C-A sequence. There are four atoms per unit cell which is marked with red rhombi in Figure 1a and each Ge (Se) atom is coordinated with three Se (Ge) atoms. The length of Se-Ge bond is 2.55 Å while the Ge-Ge bond is 2.88 Å. After a full structural relaxation, the optimized lattice parameters are $a = b = 3.763$ Å, which are accordance with previous theoretical works.[23, 42] To be an anode material, it is required a moderate adsorption to accommodate the Li. The adsorption energy per Li on the γ-GeSe surface can be calculated using the following equation

$$E_{\text{ads}} = (E_{\text{total}} - E_{\gamma-\text{GeSe}} - nE_{\text{Li}})/n \qquad (1)$$

where $E_{\text{total}}$, $E_{\gamma-\text{GeSe}}$ and $E_{\text{Li}}$ are the total energies of γ-GeSe intercalated with Li, pristine γ-GeSe and Li single atom, respectively, $n$ is the number of Li atoms. The energy of a single Li atom is calculated by putting Li atom in a non-cubic box with different dimensions of 7.5 Å × 8.0 Å × 8.9 Å. A more negative value of adsorption energy represents a stronger adsorption of Li atom on the γ-GeSe substrate. We identified the possible adsorption sites of Li on γ-GeSe nanosheet, as shown in Figure 1a. It is worth noting that the sites on the bridge will move to hollows nearby, so there are only three non-equivalent chemical environments due to high symmetry of γ-GeSe. One is on the top of Se atom and labeled as site I. Another two sites on the top of hollow centers which are directly above two inequivalent Ge atoms at different subsurface layers, labeled with II and III, respectively. The hollow II and III sites are surrounded by three triangle Se atoms. Our results show that the Li atom prefers these two hollows with $E_{\text{ads}}$ of -1.68 and -1.67 at II and III sites, respectively, much stronger than that above site I ($E_{ads} = -1.02$ eV) (seeing Figure 1b and Table 1). From Bader charge analysis, each Li atom loses 0.89-0.92 electron for these three adsorbed sites, which implies that Li atom exists in cationic state, implying a strong binding of Li atom and γ-GeSe. At the most energetically favorable site II, the distances between Li and adjacent Se atoms range from 2.58–2.60 Å shown in Figure 1c,d. During the charging/discharging process, the lattice might be changed by strain. We have examined the effect of strain on the adsorption energy, and found that with the dilation of the lattice stronger adsorptions occur albeit negligible changes of the charge transfer as shown in Figure S1. We also compared the adsorption energy of γ-GeSe nanosheet with bulk counterpart (seeing Figure S2) at the same position in Table 1, showing $E_{ads}$ of bulk γ-GeSe is stronger than monolayer case and $E_{ads}$ fall in the range from -1.84 to -2.26 eV.

More evidence of the strong adsorption of Li on γ-GeSe monolayer is further supported by charge density difference defined as:

$$\Delta\rho = \rho_{total} - \rho_{\gamma-GeSe} - \rho_{Li} \qquad (2)$$

where $\rho_{total}$, $\rho_{\gamma-GeSe}$ and $\rho_{Li}$ represent the total charge densities of system, γ-GeSe without Li and Li atom, respectively. Figure 1e,f shows the electrons accumulate between Li and adjacent Se atoms, indictive of the strong charge transfer and interaction of Li atom and substrate, which is in agreement with Bader charge analysis. Therefore, in the following discussions we focused on the most preferential site II, i.e., the hollow site on the top of Ge sublattice.

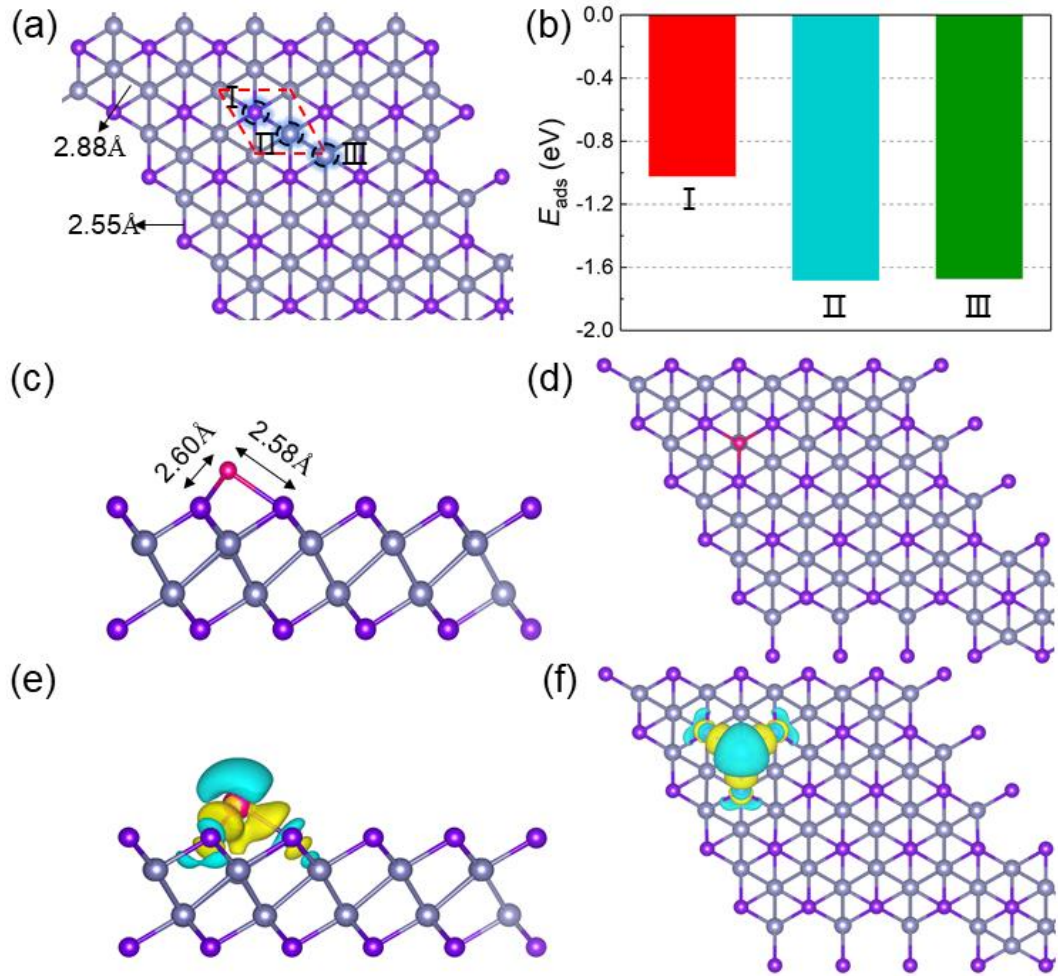

**Figure 1.** (a) Groundstate structure of γ-GeSe and three possible loading sites for Li; (b) The adsorption energies for three loading sites; The optimized structure of Li-absorbed γ-GeSe at the most energetically favorable site II: (c) side view and (d) top view; (e) The top and (f) side view of charge density difference which demonstrate the interaction of Li and γ-GeSe. The yellow (blue) color represents charge accumulation (depletion). The iso-surface level is set to 0.0015 e Å$^{-3}$. The purple, grey, and red balls represent Se, Ge and Li atoms, respectively.

**Table 1. Adsorption energies $E_{ads}$ (eV) and Bader charge (e) of Li at different sites on the γ-GeSe nanosheet and bulk phase.**

| Site Number | | I | II | III |
|---|---|---|---|---|
| Nanosheet | $E_{ads}$ | -1.02 | -1.68 | -1.67 |
| | Bader charge | +0.92 | +0.89 | +0.89 |
| Bulk | $E_{ads}$ | -1.84 | -2.26 | -1.89 |
| | Bader charge | +0.84 | +0.87 | +0.85 |

**The stability of γ-GeSe intercalated with Li.**

The synthesized γ-GeSe by a CVD process[43] exists in the bulk phase. To assess the stability of γ-GeSe nanosheet, the phonon dispersion is calculated using large enough supercell (7 × 7 × 1). The resulting phonon dispersion spectra (seeing Figure S3) without imaginary frequency validates the dynamical stability of monolayer γ-GeSe. Then, to explore the thermal stability of γ-GeSe with Li intercalation, the AIMD simulations at room temperature (*T* = 300 K) were carried out. As first, the system is thermalized at constant temperature intervals for a duration of 2 ps. Then the system is simulated under NPT ensemble using Nose$^{-'}$Hoover thermostat.[53] The simulation process ran for 10 ps with a timestep of 2 fs. As shown from Figure 2a, there is no significant distortion of the system after AIMD simulations. The temperature and energy of Li-adsorbed γ-GeSe with a time scale of 10 ps at 300 K are plotted in Figure 2b and c. The robust fluctuations of temperature and energy during AIMD simulations indicates a good stability of γ-GeSe with Li intercalation. The distances between Li and adjacent Se atoms are in the range of 2.52-2.61 Å without obvious changes compared to initial structure. Apparently, the charge donation from Li does not trigger a local change of breaking of Ge-Se bond nor a non-local lattice distortion over large scale. Our results indicate that, as a new polymorph of GeSe, the Li can have a good structural compatibility with γ-GeSe which allows potential chemical exfoliation of few layer GeSe via Li intercalation.

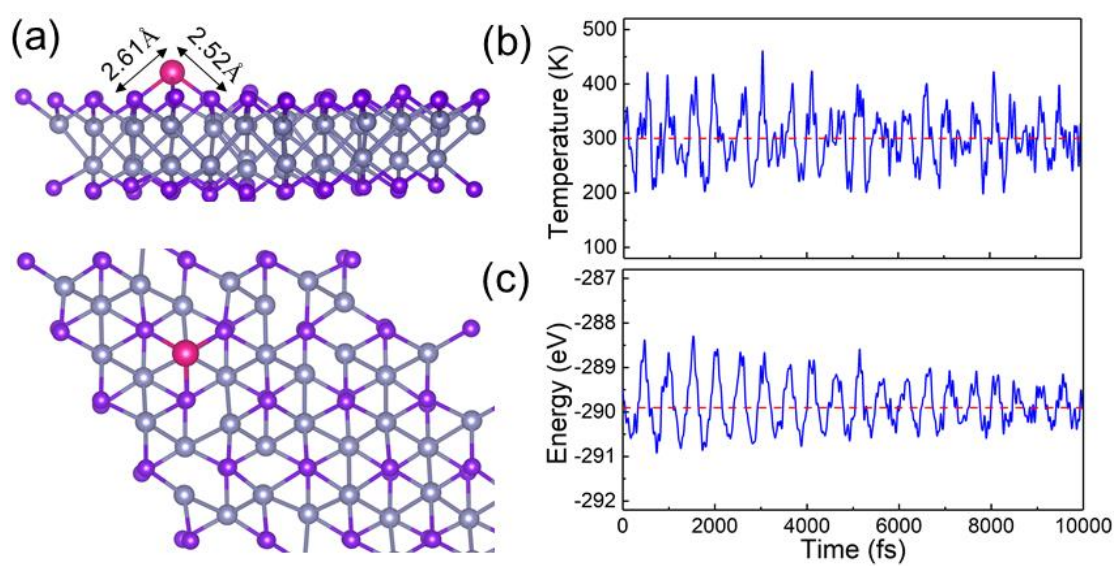

**Figure 2.** (a) Snapshot of top and side atomic structure of Li-absorbed γ-GeSe after AIMD simulations with the timescale 10 ps; Fluctuations of (b) temperature and (c) energy.

**Kinetics of Li on γ-GeSe.**

As an anode for LIB, the kinetics of Li diffusion on the γ-GeSe surface is a dominant factor for charge/discharge rate of metal-ion battery. A low diffusion barrier implies a high charge/discharge rate in LIBs. For γ-GeSe, three are three possible pathways where Li transfers from an initial hollow II site to its neighboring equivalent hollow II site (seeing Figure 3a): P1 (P2) with the Li atom diffusing via an intermediate site III (neighboring Se atom, site I), and P3 with a direct linear interpolated replica. The diffusion energy barriers for each pathway were calculated via CI-NEB to be 0.21, 0.65 and 1.92 eV along the P1, P2 and P3 pathways, respectively, which is shown in Figure 3b. Hence, the P1 route via hopping along the hollow sites is the most favorable pathway for Li diffusion, whose barrier is the lowest with 0.21 eV among three pathways.

We also calculated the diffusion behavior of Li across the vdW gap of the bulk γ-GeSe. For bulk γ-GeSe, the AB'-stacking crystal structure is used, which is the most energetically favorable sequence.[23, 42, 43] The AB'-stacking order is shown in Figure S2 and the stable adsorption sites are similar to that on γ-GeSe nanosheet. The schematic diagram of diffusion pathways and energy profiles were presented in Figure S4. It is revealed that the diffusion energy barriers for each pathway in bulk γ-GeSe were 0.52, 0.55 and 2.06 along the P1, P2 and P3 pathways, respectively. Therefore, the Li diffusing across in the bulk experiences higher barriers (nearly doubles) than above the monolayer surface. This is reasonable as normally a stronger adsorption will have a higher diffusing barrier as the adsorption energy ($E_{\text{ads}}$ of the most energetically favorable site is -2.26 eV) in bulk γ-GeSe is stronger than that of nanosheet (seeing Table 1). During the charging/discharging process, the electric field tends to have minor effect on the activation of Li species as shown in Figure S5.

Further, the diffusion barrier of Li above monolayer γ-GeSe nanosheet is superior to that of ψ-graphene (0.31 eV),[56] borophene (0.60 eV)[57] and MoN2 (0.78 eV),[58] and comparable to that of VS$_2$ (0.22 eV)[59] and CoS$_2$ (0.22 eV).[60] For quantitatively evaluating the diffusion rate, we calculated the diffusion constant (*D*), which can be obtained using Arrhenius equation:

$$D \sim \exp\left(\frac{-E_a}{k_B T}\right) \quad (3)$$

where $E_a$ is activation energy, that is diffusion barrier, $k_B$ is Boltzmann's constant and *T* is the Kelvin temperature. According to Equation 3, the diffusion rate of Li on the γ-GeSe surface is about 32 and $2.2 \times 10^6$ times faster than ψ-graphene[56] and borophene,[57] respectively. Therefore, a high-rate capability is expected to achieve in the γ-GeSe nanosheet for Li-ion battery application.

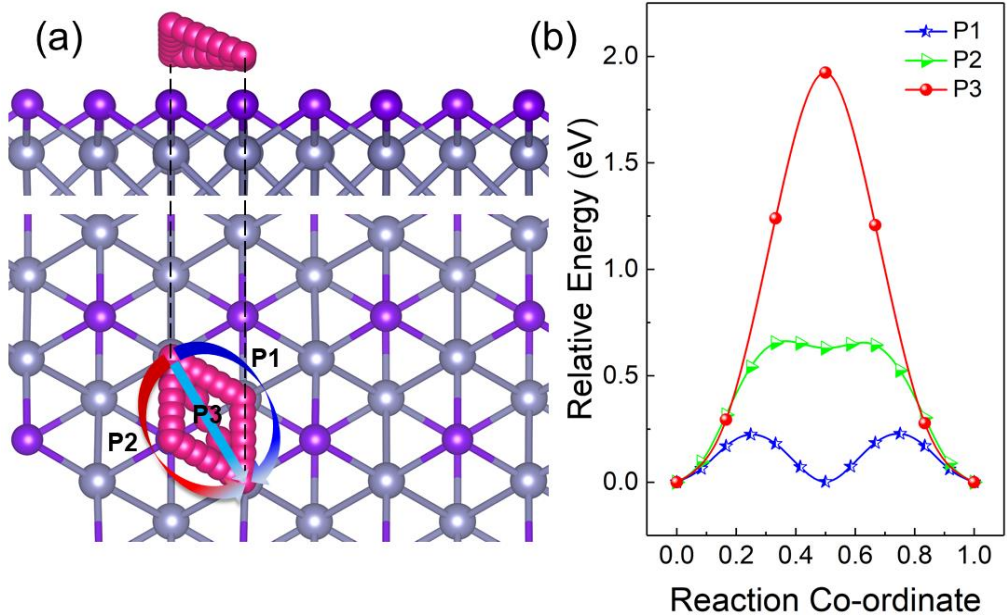

**Figure 3.** (a) The schematic diagram of diffusion of Li along three pathways: P1, P2 and P3 on the γ-GeSe nanosheet; (b) The energy barrier of P1, P2 and P3 pathways, respectively.

**The coverage effect of Li concentration and open-circuit-voltage profile.**

Next, the coverage effect of Li adsorption and average voltage were discussed. Different stoichiometry of Li$_x$GeSe ($x$ = 0.063, 0.094, 0.125 and 0.250) were selected as shown in Figure 4a-d. We tested the most energetically favorable configuration of Li$_{0.063}$GeSe where two Li atoms locate in non-equivalent position. As shown in (Figure S6), the total energy of Li$_{0.063}$GeSe where two Li atoms are in two nearest hollow II (-292.71 eV) is higher than that of where two Li atoms are in two hollow II with a gap hollow II in the middle (-292.81 eV). We therefore selected nonadjacent hollow II configurations to immobilize Li as possible for Li$_x$GeSe with higher stoichiometry. With the increasing concentration of Li insertion, the binding force between Li and γ-GeSe gradually decreases (Table 2 and Figure 4e) with a linear tendency due to the Coulomb repulsion among the Li species. The structure of γ-GeSe can be maintained stable at $x$ = 0.25, suggesting its good storage capacity.

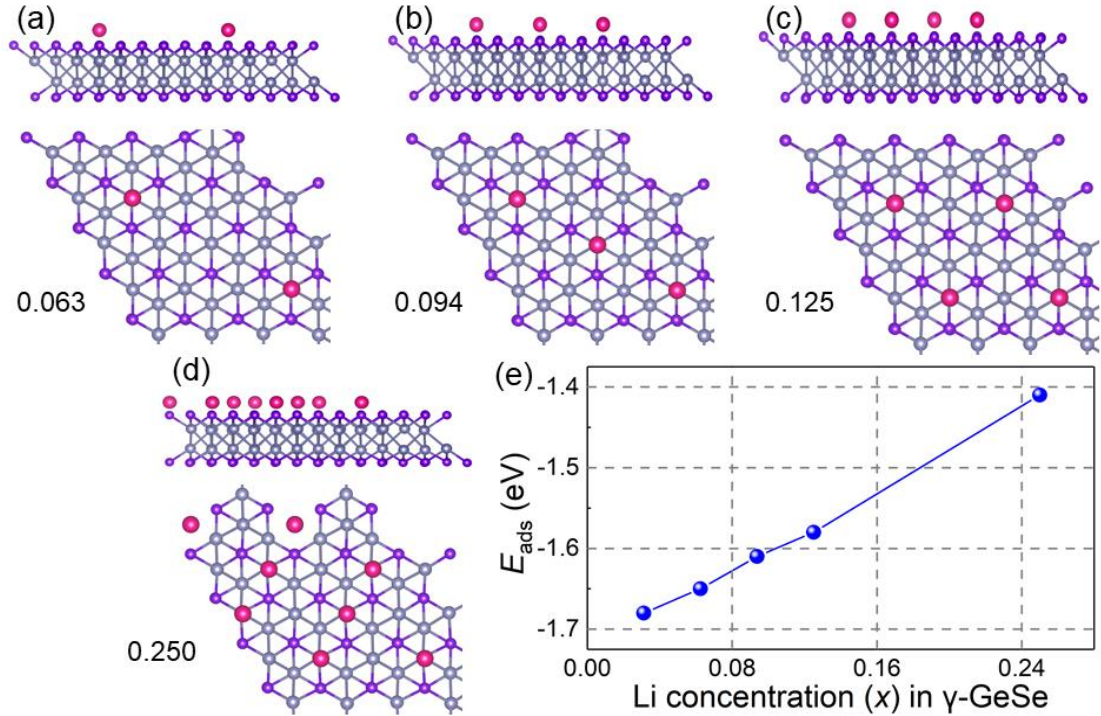

**Figure 4.** Top and side views of Li$_x$GeSe with the Li coverage of (a) 0.063, (b) 0.094, (c) 0.125 and (d) 0.250, respectively; (e) The variations of adsorbed energies with the increasing Li concentration.

**Table 2.** Adsorption energies $E_{ads}$ (eV), OCV (V) of Li and at different stoichiometry of Li$_x$GeSe.

| Stoichiometry of Li$_x$GeSe | 0.031 | 0.063 | 0.094 | 0.125 | 0.250 |
|---|---|---|---|---|---|
| $E_{ads}$ | -1.68 | -1.65 | -1.61 | -1.58 | -1.41 |
| Voltage | 0.071 | 0.065 | 0.027 | 0.027 | 0.015 |
| Volume expansion | 0.68% | 1.32% | 1.73% | 2.34% | 2.72% |

Open-circuit-voltage (OCV) is used to measure the driven potential provided by the rechargeable LIBs. The charge/discharge process of γ-GeSe for Li-ion battery obeys the half-cell reaction versus Li/Li$^+$:

$$(x_2 - x_1)\text{Li}^+ + (x_2 - x_1)e^- + \text{Li}_{x_1}\text{GeSe} \leftrightarrow \text{Li}_{x_2}\text{GeSe} \qquad (4)$$

On the basis of above reaction, the average OCV can be estimated of Li$_x$GeSe with the concentration range of $x_1 \leqslant x \leqslant x_2$ (neglect the entropy and volume effects) via:[54, 55]

$$\text{OCV} \approx \frac{E_{\text{Li}_{x_1}\text{GeSe}} - E_{\text{Li}_{x_2}\text{GeSe}} + (x_2 - x_1)E_{\text{Li}}(bulk)}{(x_2 - x_1)e} \qquad (5)$$

where $E_{\text{Li}_{x_1}\text{GeSe}}$ and $E_{\text{Li}_{x_2}\text{GeSe}}$ are the total energies of Li$_{x1}$GeSe, Li$_{x2}$GeSe, and the $E_{\text{Li}}(bulk)$ is the energy of per Li atom in the body-centered cubic lattice,[54] which is different from Equation 1. The average voltage versus Li concentration is depicted in Figure 5. It is shows that the OCV varies in the range of 0.071-0.015 V, and there is an obvious drop from 0.065 V to 0.027 V when $x$ reaches to 0.094. This drop could be caused by the promoted electrostatic repulsive interactions with the uptake of more Li atoms. When the third Li atom approaches the adsorption site, symmetrical repulsion of two other Li atoms will trigger the rise of the energy of the system. Unlike the cathode, a low OCV is beneficial for the materials as anodes (for primary battery, the electromotive force $E$ = OCV$_\text{cathode}$ - OCV$_\text{anode}$). Therefore, the OCV of commercial anode materials close to zero is desired and here the predicted voltage of γ-GeSe are comparable to commercial graphite (0.10 V). [61] Furthermore, the volume expansion of this system is smaller than 3% in the charging and discharging processes (seeing Table 2). Thus, γ-GeSe nanosheet can provide a satisfying charging voltage and minor volume variation, indicative of a promising anode material of γ-GeSe for Li-ion battery applications.

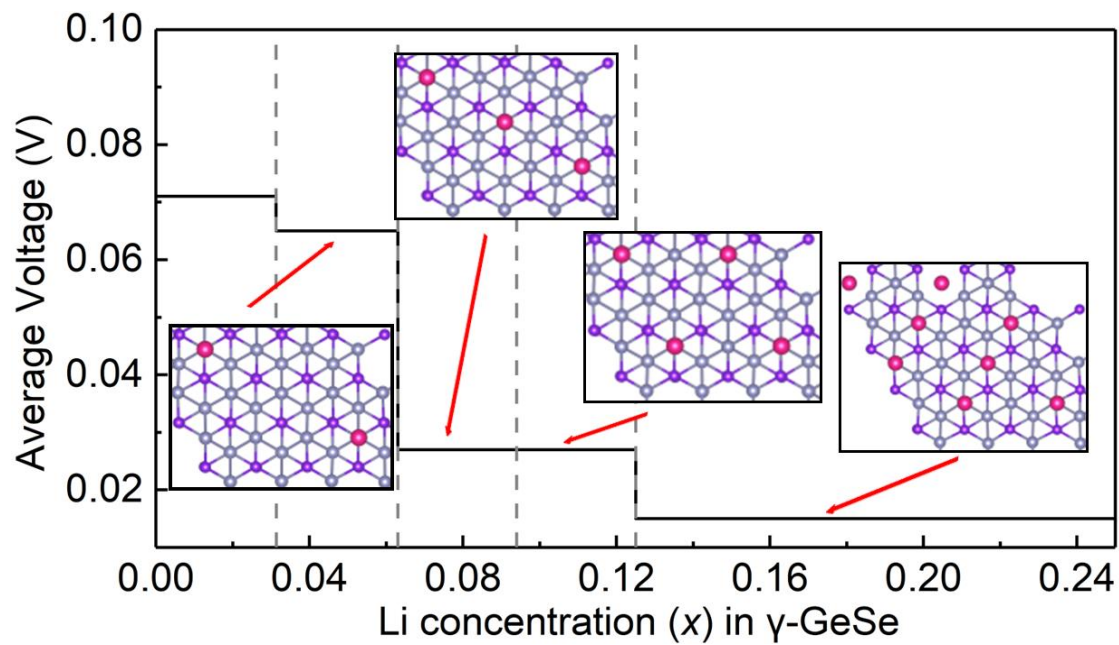

**Figure 5.** The calculated open-circuit voltage (OCV) of γ-GeSe with respect to Li concentration from 0.031 to 0.25.

**Electronic property and storage capacity of γ-GeSe coupled with Li.**

To get a physical insight for anode material of γ-GeSe with respect to LIB performance, the electronic density of states (DOS) and band structure along the high-symmetry points (K-Γ-M-K) in the first-Brillouin zone were carried out. As shown in Figure 6a, the monolayer γ-GeSe nanosheet is shows a semiconducting characteristic with an indirect bandgap of 0.59 eV. When Li is absorbed on the γ-GeSe, several bands across the Fermi level (seeing Figure 6b) can be observed compared with pristine γ-GeSe, implying the metallic nature of Li-absorbed γ-GeSe at equilibrium. As shown in the PDOS plot in Figure 6c and d, the intercalated Li causes negligible changes of p-orbitals of Ge and Se except the upward shift of the Fermi level to the conduction band. We would like to demonstrate that GGA-PBE functionals underestimate the electronic properties of γ-GeSe and Li-absorbed system, whereas the hybrid functionals (such as HSE06) are able to acquire more accurate band gaps. However, GGA-PBE is enough to estimate the metallic nature of metal-doped system[8,55] and HSE06 is computationally expensive, so vdW-correlated GGA-PBE functionals are chosen to calculate the electronic properties in this work.

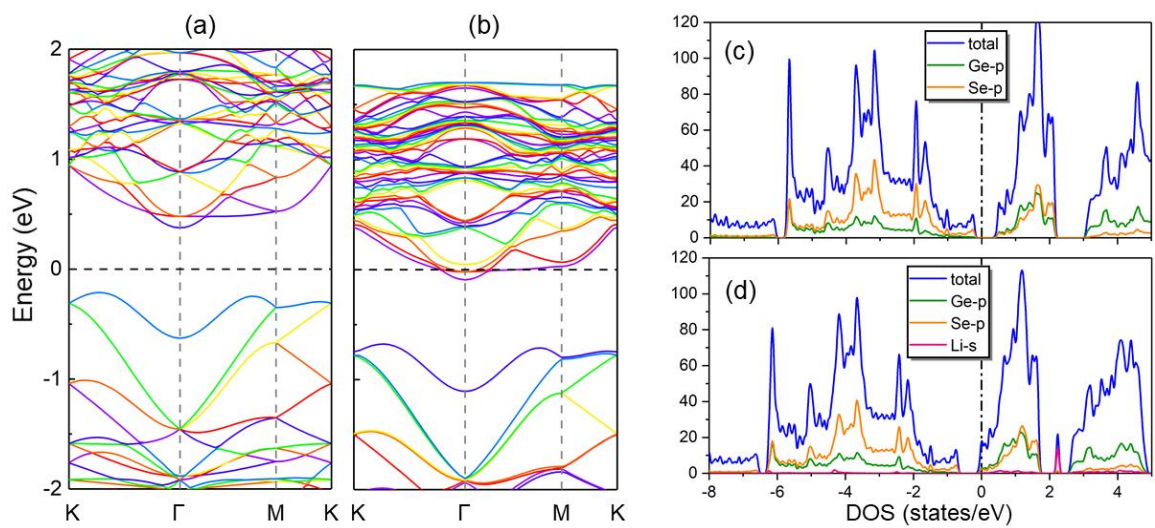

**Figure 6.** Band structure and electronic DOS of (a) pristine γ-GeSe and (b) Li-absorbed γ-GeSe, respectively. Noted the magnitude of Li-s state is enlarged by a factor of 50. The Fermi level represented by dashed line is set to zero.

Band decomposed charge densities (Fermi level ± 0.3 eV) around band edges are calculated and shown in Figure 7. The distribution of states near the Fermi level are found to mostly accumulate on around hollow II for $Li_{0.125}GeSe$ shown in Figure 7a and coincided the conducting channel that supports the diffusion barrier calculation. For the slightly increased concentration of Li to 0.25 ($Li_{0.25}GeSe$), the distribution of conducting electrons still displays directional dependence (seeing Figure 7b). Our calculation indicates the real space occupancies of additional electrons and naked Li cations are both at the hollow sites.

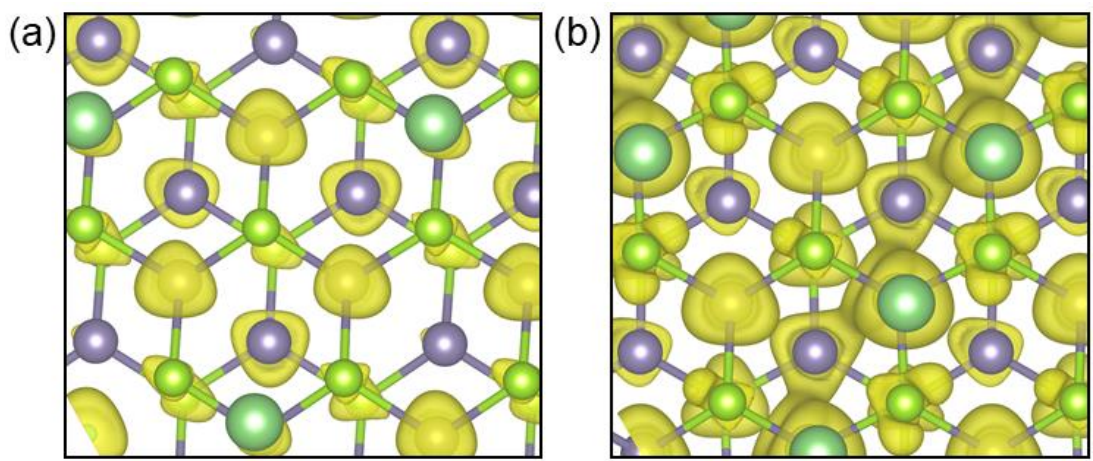

**Figure 7.** Electron density distributions of local regions around the Fermi level of (a) $Li_{0.125}GeSe$ and (b) $Li_{0.25}GeSe$. The iso-surface level is set to 0.0035 e Å$^{-3}$. The pale green, blue cyan, and dark green balls represent Se, Ge and Li atoms, respectively.

In order to check out the storage capacity of γ-GeSe nanosheet, the theoretical storage capacity (mAh g$^{-1}$) can be evaluated as:

$$C = \frac{xF}{M} \quad (6)$$

Here, $x$ is the number of absorbed Li ions, $F$ is the Faraday constant (26801 mAh g$^{-1}$), and $M$ is the molecular mass of the given material. When $x$ reaches to 3 of Li$_x$GeSe, the value of OCV is 0.078 V. The high OCV of Li$_3$GeSe is caused by the structural distortion. However, there is no bond breaking in γ-GeSe. The optimized structures with different Li content ($x$ = 1, 2, 3) are shown in Figure. S7. The theoretical storage capacity of γ-GeSe is found to be over 530.36 mAh g$^{-1}$, which is a reasonably good capacity for LIBs application as an anode material. Our predicted value of capacity for γ-GeSe is larger than those of some other 2D materials[17, 18, 55,57].

## Conclusions

In summary, the new semi-metallic polymorph of γ-GeSe which shares some features with graphene such as high conductivity and a hexagonal lattice for LIBs applications. For the first time we have performed systematic first-principles calculations to evaluate the energetics and kinetics of Li across the γ-GeSe layer. We found the Li atoms undergo a small barrier of ~0.21 eV above monolayer while this values doubles for across bulk γ-GeSe. There is no clustering tendency of Li atoms above the layer. Our work shows the intermediate level of adsorption and fast diffusion suggests a possible chemical exfoliation of γ-GeSe via Li intercalation. This high mobility of Li also allows a potential LIBs application. We hope that this work will promote further interests on the applications of γ-phase group-IV monochalcogenides for theoretical and experimental research.

*Acknowledgments*

This work is supported by the Natural Science Foundation of China (Grant 22022309) and Natural Science Foundation of Guangdong Province, China (2021A1515010024), the University of Macau (SRG2019-00179-IAPME, MYRG2020-00075-IAPME) and the Science and Technology Development Fund from Macau SAR (FDCT-0163/2019/A3). This work was performed in part at the High-Performance

Computing Cluster (HPCC) which is supported by Information and Communication Technology Office (ICTO) of the University of Macau.

*Conflict of Interest*

The authors declare no conflict of interest.